\documentclass[a4paper,onecolumn,11pt]{quantumarticle}
\pdfoutput=1
\usepackage[utf8]{inputenc}
\usepackage[english]{babel}
\usepackage[T1]{fontenc}
\usepackage{amsmath}
\usepackage{hyperref}
\usepackage{comment}

\usepackage{tikz}
\usetikzlibrary{quantikz}
\usetikzlibrary{decorations.markings}
\usepackage{lipsum}

\newcommand{\eps}{\varepsilon}
\newcommand{\tr}{\mathrm{Tr}}

\newtheorem{definition}{Definition}
\newtheorem{theorem}{Theorem}

\newtheorem{corollary}{Corollary}

\begin{document}

\title{Alpha-bit state merging}

\author{Jessica Yeh}
\affiliation{Leinweber Institute for Theoretical Physics, Stanford University, Stanford, CA 94305}
%\email{ycy@stanford.edu}

% \orcid{...}
\author{Jinzhao Wang}
%\affiliation{Stanford University}
%\email{jinzhao@stanford.edu}
\author{Patrick Hayden}
%\email{phayden@stanford.edu}
% \homepage{http://quantum-journal.org}
% \orcid{...}

\maketitle

\begin{abstract}
State merging is a fundamental protocol in quantum information theory that generalizes quantum teleportation. Traditionally, it is achieved by local operations on shared entanglement and classical communication. In this work, we study state merging done with \emph{$\alpha$-bits}, a versatile quantum communication resource weaker than qubits. %We find that the resource inequality of state merging simplifies with $\alpha$-bits. 
We study $\alpha$-bit state merging with and without catalytic entanglement, and we find a potential gap between the rates of $\alpha$-bits consumed. In light of our result, we discuss how to interpret entanglement wedge reconstruction in AdS/CFT in terms of $\alpha$-bit state merging.
\end{abstract}

\section{Introduction}
Quantum teleportation is a fundamental quantum information-processing protocol that transmits a qubit by sending two classical bits while consuming one EPR pair \cite{bennet1993teleportation}. We can summarize its resource consumption using a resource inequality~\cite{DevHarWin2008resource}. For instance, a teleportation protocol can be written as 
\begin{equation}
    1 [qq] + 2 [c\rightarrow c] \geq 1 [q\rightarrow q]
\end{equation}
where $[qq]$ represents sharing an EPR pair, $[c\rightarrow c]$ represents classical bits communication, and $[q\rightarrow q]$ represents quantum bits communication.

\textit{State merging}~\cite{HOW2005} is a generalization of teleportation where the sender (Alice) and the receiver (Bob) share a state $\rho_{AB}$ and the goal is to have Alice transfer her share to Bob using classical communication and entanglement. An alternative and equivalent way of formulating the task is by considering a purification $|\psi\rangle_{ABR}$, shared by Alice ($A$), Bob ($B$), and a referee ($R$). That is, Alice and Bob collectively hold the purification of the referee's state. The goal of state merging is then to transfer all the purification of $R$ to $B$ via LOCC between $A$ and $B$ while the state at $R$ remains unchanged~\cite{HOW2005}. In the case when $B$ is trivial, this reduces to standard teleportation. 

Remarkably, in~\cite{HOW2005}, it is shown that the rate of entanglement required to achieve state merging in the asymptotic setting is the conditional entropy $H(A|B)_\psi$, operationally realizing the idea that conditional entropy should quantify Bob's ignorance of Alice's state. In addition, the protocol requires sending classical bits at rate $I(A:R)_\psi$. In the language of resource inequalities, we can write
\begin{equation}
\label{eqn:HOW state merging}
    \langle \psi_{ABR} \rangle + H(A|B)_\psi [qq]  + I(A:R)_\psi [c \rightarrow c] \geq \langle \psi_{A'BR} \rangle.
\end{equation}
Here we use $\psi_{A'BR}$ to denote the final state where Bob holds the purification $A'B$ of $R$ and $A'\equiv A$. 
The inequality means that by consuming entanglement at rate $H(A|B)_\psi$ using a rate $I(A:R)_\psi$ of classical communication from Alice to Bob, we can convert copies of the initial state $\psi_{ABR}$ into copies of the merged state $\psi_{A'BR}$.

Since teleportation amounts to simulating quantum communication using entanglement and classical communication, one can also reformulate state merging by only allowing quantum communication between Alice and Bob. This is known as the \textit{mother protocol}, or the \textit{fully quantum Slepian-Wolf protocol}, which has the following resource inequality~\cite{ADHW2009mother,DatHsi2011mother}:
\begin{equation}
\label{eqn:fqsw}
\langle \psi_{ABR}\rangle + \frac{1}{2}I(A:R)_\psi [q\rightarrow q] \geq  \langle \psi_{A'BR} \rangle + \frac{1}{2}I(A:B)_\psi [qq]\ .
\end{equation}
That is, we achieve state merging by directly sending qubits at the rate $I(A:R)_\psi/2$. The $I(A:B)_\psi/2$ on the right-hand side means that we gain an extra entanglement at the rate $I(A:B)_\psi/2$ at the end of the protocol. From this primitive protocol, one can deduce various variants of quantum state transfer.

In this paper, we investigate the task of state merging with a weaker quantum resource called the \textit{$\alpha$-bits}~\cite{HP2017,HP2018}, where $0\le\alpha\le 1$. Intuitively, one should think of an $\alpha$-bit as an $\alpha$-fraction of a qubit in the sense that one can only retrieve faithfully an $\alpha$-fraction information through error correction in a noisy qubit. 

Conceptually, $\alpha$-bits encapsulate the task of \textit{approximate universal subspace error correction}, which differs from the usual notion of approximate quantum error correction in two important ways. First, it do not require correcting the whole space but only a subspace, say of dimension $k$. Moreover, it requires a decoding scheme to exist for \textit{every} subspace of dimension $k$. Roughly speaking, if Alice sends her state to Bob with $\alpha$-bits, then Bob can approximately recover this state as long as it lies in a subspace of dimension no larger than $k=d_A^{\alpha}$.  We provide a self-contained review of $\alpha$-bits in Sec.~\ref{sec:alpha-bits}. 

As shown in~\cite{HP2017}, $\alpha$-bits can interpolate between various resources, and hence help tighten, generalize, and simplify resource inequalities of various tasks. We will show that is also the case for state merging with $\alpha$-bits. Using $\alpha$-bits leads to a particularly simple form of the resource inequality of state merging. For $\alpha$-bits with $\alpha=\frac{H(A|B)_\psi}{H(A)_\psi}$, one has
\begin{equation}
  \langle \psi_{ABR}\rangle + H(A)_\psi [\alpha] \stackrel{(c)}{\geq} \langle \psi_{A'BR} \rangle\ ,
\end{equation}
where the symbol $(c)$ means that this protocol uses catalytic entanglement; the transformation is achieved by first borrowing some entanglement and then returning the same amount (up to sublinear loss) at the end. 

Compared to the standard state merging  of \eqref{eqn:HOW state merging} and the mother protocol \eqref{eqn:fqsw}, catalytic $\alpha$-bit state merging does not require additional input and does not leave any leftover resources. This is the cleanest form of state merging. This simplicity makes the $\alpha$-bit an important resource to study in the context of state merging. It also provides an operational meaning for $\alpha$ bits: Bob's ignorance of Alice's state is captured by the fraction $\alpha=\frac{H(A|B)_\psi}{H(A)_\psi}$.

Besides the information-theoretic considerations, $\alpha$-bit state merging also has a role to play in field of quantum gravity. Specifically, it has been shown that $\alpha$-bits arise naturally as an available resource in the context of the AdS/CFT correspondence~\cite{HP2018}. Moreover, state merging with $\alpha$-bits has been proposed as a way to realize an important component of the AdS/CFT dictionary called \textit{entanglement wedge reconstruction}. This proposal is supported by the matching of information resource consumption for one-shot state merging and the criteria for entanglement wedge reconstruction~\cite{AP2020}. One caveat to the protocol proposed by~\cite{AP2020} is that it is only shown to be be realizable using catalytic entanglement. Whether catalytic entanglement is actually available in AdS/CFT is not addressed. 

The use of catalytic entanglement has been studied extensively in other quantum information theory settings, and has been found to be indispensable for certain tasks. Readers can consult~\cite{lipka2024catalysis} for a recent review.
There are many situations that demonstrate an advantage when catalytic entanglement is present. For example, for the task of LOCC entanglement transformation, Nielsen showed that $|\psi\rangle_{AB}$ can be transformed into $|\phi\rangle_{AB}$ if and only if $\phi_{AB}$ majorizes $\psi_{AB}$~\cite{nielson1999ent-transformation}. Using that condition, one can prove that there are cases where it is possible to convert $|\psi\rangle_{AB}$ into $|\phi\rangle_{AB}$ only in the presence of catalytic entanglement. That is, $|\psi\rangle_{AB} \otimes |\tau\rangle_{AB} \rightarrow |\phi\rangle_{AB} \otimes |\tau\rangle_{AB} $ is possible, where $|\tau\rangle_{AB}$ is some entangled state but $|\psi\rangle_{AB} \rightarrow |\phi\rangle_{AB}$ is not~\cite{jonathan1999catalytic-entanglement}. 

On the other hand, there are also examples where the presence of catalytic entanglement does not provide further advantage. For instance, the quantum capacity of a quantum channel is given by the regularized coherent information of the channel, and this can be achieved with or without catalytic entanglement \cite{devetak2005channel,hayden2008decoupling,Wildebook}. 

It is hence crucial to figure out how important catalysis is for $\alpha$-bit state merging. In this work, we systematically study $\alpha$-bit state merging in more detail in both the catalytic and non-catalytic settings, finding that one generally needs strictly more $\alpha$-bits to achieve state merging in the absence of catalytic entanglement. That is our main result. In the setting of entanglement wedge reconstruction, one may not have access to a large enough number of catalytic ebits, it is necessary to carefully revisit the state merging / AdS/CFT connection in light of our results.

This paper is organized as follows. In Sec.~\ref{sec:background}, we review necessary prerequisites about $\alpha$-bits and some resource inequalities. In Sec.~\ref{sec:catalytic}, we rephrase $\alpha$-bit state merging in~\cite{AP2020} and show that it is optimal. In Sec.~\ref{sec:non-catalytic}, we show the achievability and optimality of the non-catalytic version of $\alpha$-bit state merging. Finally, in Sec.~\ref{sec:holography}, we discuss the implications of our results in AdS/CFT.

\subsection{Notation}
\label{sec:notation}
We use $S(A)$ to denote all density matrices in $\mathcal{H}_A$ and $\phi$ to denote the density matrix of the pure state $|\phi\rangle$.

We write $[\text{LHS}] \geq [\text{RHS}]$ to indicate that there is a protocol that turns the resources on the LHS into the resources on the RHS. For example, a teleportation protocol can be written as 
\begin{equation}
    1 [qq] + 2 [c\rightarrow c] \geq 1 [q\rightarrow q]\ ,
\end{equation}
where we use $[qq]$ for ebits (shared Bell pairs between the sender and the receiver),  $[c\rightarrow c]$ for cbits (the sender can send classical bits to the receiver), $[q\rightarrow q]$ for qubits (the sender can send  qubits to the receiver). In the following, we also have $[q\rightarrow qq]$ for cobits,  $[\alpha]$ for $\alpha$-bits, and $\langle \psi \rangle$ for having the state $\psi$ as a resource. For state merging, we use $\langle \psi_{ABR} \rangle$ to indicate the initial state before merging, and $\langle \psi_{A'BR} \rangle$ to indicate the merged state where $A'$ belongs to Bob. 

Since we are interested in the discrepancy between the catalytic and non-catalytic cases, we indicate a resource inequality achieved by a catalytic protocol with $\stackrel{(c)}{\geq}$. If the protocol can be achieved non-catalytically then we denote it with the usual $\geq$.  Resource inequalities are defined \textit{asymptotically} so, unless otherwise noted, the coefficients should be interpreted as rates in the limit that many copies of the resources described on the left are converted into those on the right. The formalism is developed rigorously in~\cite{DevHarWin2008resource}.

\section{Quantum communication with $\alpha$-bits}
\label{sec:background}
\subsection{$\alpha$-bits}
\label{sec:alpha-bits}
The $\alpha$-bit concept is first introduced in~\cite{HP2017} and is discussed in the context of black holes in~\cite{HP2018}. Like qubits, ebits, and cbits, $\alpha$-bits are a type of non-local resource. Noiseless classical communication can be viewed as a degradation of quantum communication in which only the information transmitted in a fixed basis is properly transmitted. $\alpha$-bits likewise correspond to a restriction on the type of information that is properly transmitted but the restriction takes the form of the size of the subspace the channel will preserve.
Before defining $\alpha$-\textit{bits} we first define $\alpha$-dits for technical convenience.
\begin{definition}($\alpha$-dit)
    Let $d_A$ be the dimension of the system $A$. We say $A$ is sent to $B$ as an $\alpha$-dit with error $\epsilon$ through a channel $\mathcal{N}^{A\rightarrow B}$ if for all subspaces $\Tilde{A}$ of $A$ with $|\Tilde{A}| \leq |A|^{\alpha} + 1$,
    there exists a decoding channel $\mathcal{\tilde{D}}^{B\rightarrow \tilde{A}}$ such that  
    $ ||(\mathcal{\tilde{D}} \circ \mathcal{N} \circ \mathcal{E}\otimes \operatorname{Id}_{R})\phi^{AR}-\phi^{AR}||_1 \leq \epsilon $ for all $|\phi\rangle$ in $\tilde{A}R$. The channel $\mathcal{E}$ is the encoding channel used by $A$.
\end{definition} 

\begin{figure}
    \begin{tikzpicture}[scale=1.1]
        \draw (0,0) -- (1,2);
        \draw (1,2) -- (6.5,2); % R line
        \draw (0,0) -- (1,-1);
        \draw (1,-1) -- (1.5,-1); %\tilde{A} line
        \draw (1.5,-1) -- (1.5,-0.5) -- (2.5,-0.5) -- (2.5, -1.5) -- (1.5,-1.5) -- (1.5,-1); %V_e box
        \draw (2.5,-0.6) -- (3,-0.6); %A line
        \draw (2.5,-1.4) -- (6.5,-1.4); % E line
        \draw (3,-0.6) -- (3,-0.1) -- (4,-0.1) -- (4,-1.1) -- (3,-1.1) -- (3,-0.6); % U_N box
        \draw (4,-1) -- (7,-1); % E' line
        \draw (4,-0.6) -- (5,0.2); % B line
        \draw (5,0.2) -- (5,0.7) -- (6,0.7) -- (6,-0.3) -- (5,-0.3) -- (5,0.2); %V_D box
        \draw (6,0.4) -- (6.5,0.4); %second \tilde{A} line
        \draw (6.5,0.4) -- (7,1.2);
        \draw (6.5, 2) -- (7,1.2);
        \draw (6,-0.2) -- (6.5,-0.2); % E'' line
        \draw (6.5,-0.2) -- (7,-1);
        \draw (6.5,-1.4) -- (7,-1);
        \node[above] at (1.2,2) {$R$};
        \node[above] at (1.2,-1) {$\tilde{A}$};
        \node[above] at (27,-0.6) {$A$};
        \node[above] at (4.3,-0.3) {$B$};
        \node[right] at (1.7,-1) {$V_{\varepsilon}$};
        \node[right] at (3.2,-0.6) {$U_{\mathcal{N}}$};
        \node[right] at (5.2,0.2) {$V_D$};
        \node[above] at (6.2,0.4) {$\tilde{A}$};
        % RHS graph
        \node[right] at (7,0) {$=$};
        \draw (8,0) -- (9,2);
        \draw (9,2) -- (13,2); %R line
        \draw (8,0) -- (9,-1);
        \draw (9,-1) -- (9.5,-1); %\tilde{A} line
        \draw[thick] (9.5,-1) -- (12,0.5); %\alpha line
        \draw (12,0.5) -- (13,0.5); %second \tilde{A} line
        \draw[->,thick] (9.5,-1) -- (11,-0.1);
        \draw (13,0.5) -- (13.5,1.25);
        \draw (13,2) -- (13.5,1.25);
        \node[above] at (9.3,2) {$R$};
        \node[above] at (9.3,-1) {$\tilde{A}$};
        \node[above] at (12.5,0.5) {$\tilde{A}$};
        \node[above] at (11,0) {$\alpha$};
        
    \end{tikzpicture}
    \caption{Circuit diagram for the $\alpha$-bit channel. $V_{\varepsilon}$ is the dilation of Alice's encoding channel, $U_{\mathcal{{N}}}$ is the dilation of the transmitting channel, and $V_D$ is the dilation of Bob's decoding channel. The decoding is possible if $|\tilde{A}|\leq |A|^{\alpha}$. }
    \label{fig:alpha}
\end{figure}
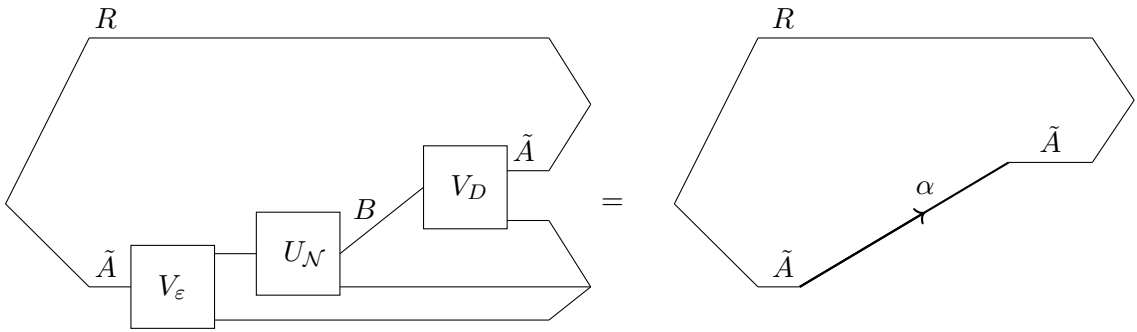

In other words, sending an $\alpha$-dit means that the receiver can decode the state if it is from some subspace of dimension less than $|A|^{\alpha}$. A circuit diagram illustrating $\alpha$-dit transmission is depicted in Fig.~\ref{fig:alpha}.

In the following, we will be mainly discussing $\alpha$-bits, which can be thought of as having $\alpha$-dits with vanishing error for sufficiently large $d$. Concretely, we will only be discussing a quantum channel's capability to transmit $\alpha$-bits. This is characterized by the $\alpha$-bit capacity, and that is how many $\alpha$-bits a channel can send per use.
\begin{definition} ($\alpha$-bit capacity)
    We say that we can transmit $\alpha$-bits at rate $Q$ through a channel $\mathcal{N}$ if for all $\epsilon>0$ and sufficiently large $d$ and $n$, we can transmit $\left \lceil \frac{nQ}{\log{d}}\right\rceil$ $\alpha$-dits with error $\epsilon$ using $\mathcal{N}^{\otimes n}$. The $\alpha$-bit capacity is the supremum over achievable rates.
\end{definition}

An equivalent formulation of $\alpha$-bits that will be important later is related to the \textit{subspace decoupling duality} proved in~\cite{HP2017}. Before stating the duality, let's review the standard decoupling principle in the context of approximate quantum error correcting codes. (See, for example,~\cite{preskillNotes}.) Consider a channel $\mathcal{N}^{A\rightarrow B}$ and its Stinespring dilation $V^{A \rightarrow BE}$. For an input state $\rho_A$ with a purification $|\psi\rangle_{RA}$ where $R$ is the reference of $A$, the decoupling principle says that $B$ can decode/recover $\rho_A$ if and only if $R$ and $E$ decouple after the isometry $V^{A \rightarrow BE}$. That is, let $|\phi\rangle_{RBE}=V |\psi\rangle_{RA}$ and $\sigma_{RE}$ be the reduced density matrix, then we have $\sigma_{RE} \approx \sigma_R \otimes \sigma_E$ if and only if $B$ can recover $\rho_A$.

The equivalent definition of $\alpha$-bits says something similar. Recall that we first define transmitting $\alpha$-bits as the Bob's ability to decode every subspace of $A$ with dimension less than $|A|^{\alpha}$. Again, let $V^{A\rightarrow BE}$ be the Stinespring dilation of the channel that transmits $\alpha$-bits and $R$ be a reference state that $A$ is entangled with at the beginning with the restriction that $|R|\leq |A|^{\alpha}$. The subspace decoupling duality then implies that we can transmit $\alpha$-bits from $A$ to $B$ if and only if $R$ and $E$ decouples after the isometry $V$. (See Corollary~\ref{cor:decouple_less_than_k} below.) Note that when $\alpha=1$, this reduces to the decoupling principle above because $A$ can be purified by a system with the same dimension. 

To formally state the subspace decoupling duality, we need the following definitions:
\begin{definition} (k-diamond norm)
    The k-diamond norm of a channel $\mathcal{N}^{A\rightarrow B}$ is $$||\mathcal{N}||^{(k)}_\diamond = \max_{\rho} ||(\operatorname{Id}_R \otimes \mathcal{N}) \rho^{RA}||_1$$
    where the maximization is taken over all density operators on $R \otimes A$ with $|R| \leq k$.
    The diamond norm of a channel is $$||\mathcal{N}||_\diamond = \sup_k ||\mathcal{N}||^{(k)}_\diamond \ .$$
\end{definition} 
We also have 
$$ ||\mathcal{N}||_\diamond =|| \mathcal{N}||_\diamond^{(d_A)}$$ because any $\rho_A$ can be purified with a reference of dimension $d_A$.

\begin{definition} (Forgetfulness)
    A channel $\mathcal{C}^{A \rightarrow E}$ is $k$-forgetful with error $\delta$ if 
    $$||\mathcal{C}-\mathcal{R}||^{(k)}_\diamond \leq \delta$$
    where $\mathcal{R}$ is a channel that takes all states in $S(A)$ to some fixed state $\omega$ in $S(E)$.
\end{definition}

The statement of the subspace decoupling duality is that being able to decode a subspace after sending the state through a channel is equivalent to the complementary channel being $k$-forgetful. That is,
\begin{theorem} (Subspace decoupling duality~\cite{HP2017})
\label{thm:subspace-decoupling}
    The following two statements are equivalent with some dimension-independent universal relation between $\epsilon$ and $\delta$:
    
    1. For all subspaces $\tilde{A} \subset A$ of dimension less than $k$, there exists a decoding channel $\mathcal{\tilde{D}}^{B\rightarrow \tilde{A}}$ such that $$\|\mathcal{\tilde{D}}\circ \mathcal{\tilde{N}}-\operatorname{Id}\,\|_\diamond \leq \epsilon$$  where $\mathcal{\tilde{N}}$ is the restriction of the channel $\mathcal{N}^{A\rightarrow B}$ to $\tilde{A}$.

    2. The complementary channel $\mathcal{N}^c$ is k-forgetful:
    $$\|\mathcal{N}^c-\mathcal{R}\|_\diamond^{(k)}\leq \delta$$
    where $\mathcal{R}^{A\rightarrow E}$ is a channel that takes all states in $S(A)$ to some fixed state $\omega$ in $S(E)$.
\end{theorem}

The above theorem is proved in~\cite{HP2017}, as a generalization to the weak decoupling duality theorem proved in~\cite{hayden2010weakduality}. The subspace decoupling duality then immediately implies the following corollary that we will use later.
\begin{corollary}
\label{cor:decouple_less_than_k}
    Let $\mathcal{N}$ be an $\alpha$-dit channel with error $\epsilon$ from $A$ to $B$, $R$ be the reference system of $A$, and $E$ be the environment of the channel $\mathcal{N}$. If $A$ can send an $\alpha$-dit to $B$ through $\mathcal{N}$, then any $\rho^{AR}$ decouples (up to some dimension-independent error $\delta$) after going through the channel as long as $|R| \leq |A|^{\alpha}$.
\end{corollary}

To see this, note that having an $\alpha$-dit with error $\epsilon$ implies $$||\mathcal{\tilde{D}}\circ \mathcal{\tilde{N}}-\operatorname{Id}||_\diamond \leq \epsilon$$ 
where $\mathcal{\tilde D}$ and $\mathcal{\tilde N}$ are the channels restricted to a subspace $\tilde{A} \subset A$ with dimension less than $|A|^{\alpha}$. 
From the subspace decoupling inequality and the definition of the $k$-diamond norm, we have that for any $|R| \leq |A|^{\alpha}$,
$$||\operatorname{Id}_R \otimes \mathcal{N}^c(\rho^{RA})-\operatorname{Id}_R \otimes \mathcal{R}(\rho^{RA})||_1 \leq \max_{\rho} ||\operatorname{Id}_R \otimes (\mathcal{N}^c-\mathcal{R})\rho^{RA} || = ||\mathcal{N}^c-\mathcal{R}||_\diamond^{(|A|^{\alpha})} \leq \delta\ .$$ 
Since $\mathcal{R}$ takes all states in $A$ to a fixed state $\omega$ in $S(E)$, we have (omitting $\operatorname{Id}_R$)
$$||\mathcal{N}^c (\rho^{RA})-\mathcal{R}(\rho^{RA})||_1 = ||\mathcal{N}^c (\rho^{RA}) - \rho^R \otimes \omega^E|| = ||\rho^{RE} - \rho^R \otimes \omega^E|| \leq \delta\ .$$
Therefore, $k$-forgetfulness of the complementary channel $\mathcal{N}^c$ implies decoupling between $R$ and $E$ for $R$ with dimension less than or equal to $|A|^{\alpha}$.

\subsection{Quantum protocols}
In this subsection, we describe some known protocols that will be the ingredients for $\alpha$-bit state merging.
\subsubsection{Mother protocol}
As we discussed in the introduction, the mother protocol~\cite{ADHW2009mother} can achieve state merging with quantum resources characterized by the following inequality:
\begin{equation}
\langle \psi_{ABR}\rangle + \frac{1}{2}I(A:R)_\psi [q\rightarrow q] \geq  \frac{1}{2}I(A:B)_\psi [qq]  + \langle \psi_{A'BR} \rangle\ .
\end{equation}

Decoupling is again a key ingredient in the proof. Consider the circuit diagram in Fig.~\ref{fig:non-catalytic-optimal-proof}. $B$ can achieve state merging if and only if $E$ and $R$ are decoupled at the dashed line. Thus, $A$'s goal is to send qubits to $B$ in a way that destroys its correlation with $R$. It turns out that $I(A:R)_\psi/2$ is the minimal qubit transmission rate needed, and it is also achievable by applying a Haar random unitary on $A$.

\subsubsection{Coherent dense coding/teleportation}
We define a \textit{cobit} as the ability to perform the following isometry~\cite{harrow2004cobit,Wildebook}:
\begin{equation}
    \alpha |0\rangle_A + \beta |1\rangle_A \rightarrow \alpha |0\rangle_A |0\rangle_B + \beta |1\rangle_A |1\rangle_B \ .
\end{equation}
We denote a cobit as $[q\rightarrow qq]$.

Cobits can be used to perform teleportation. The resource inequality is written as~\cite{harrow2004cobit}:
\begin{equation}
\label{eqn:coherent-dense-coding}
1 [q\rightarrow q] + 1 [qq] = 2[q\rightarrow qq] \ .
\end{equation}

When the resource inequality is written as an \emph{equality}, it means that it holds in both directions: the forward direction says that we can perform dense coding that sends cobits, and the reverse direction says that we can do teleportation with two cobits. 

\subsubsection{$\alpha$-bit dense coding/teleportation}
An analogous formula holds when we replace qubits with $\alpha$-bits, in the form of the following resource inequality~\cite{HP2017}: 
\begin{equation}
\label{eqn:alpha-dense-coding}
1 [\alpha] + 1[qq] \stackrel{(c)}{=} (1+\alpha) [q\rightarrow qq]\ .
\end{equation}
When $\alpha=1$, this reduces to \eqref{eqn:coherent-dense-coding}. From this expression, we see that $\alpha$-bits generalize qubits by enabling dense coding/teleportation at different rates.
Moreover, by setting $\alpha=0$ in \eqref{eqn:alpha-dense-coding} and using \eqref{eqn:coherent-dense-coding}, we obtain the zero-bit dense coding/teleportation
\begin{equation}
    \label{eqn:zero-bit-teleportation}
    1 [qq] + 2 [0] \stackrel{(c)}{=} 1 [q\rightarrow qq]\ .
\end{equation}
 
\subsubsection{Relating $\alpha$-bits and zero-bits}
We can obtain some useful relations between $\alpha$-bits and zero-bits by further manipulating the above equations. For example, subtracting $(1+\alpha)$ times \eqref{eqn:zero-bit-teleportation} from \eqref{eqn:alpha-dense-coding} we obtain
\begin{equation}
    1 [\alpha] \stackrel{(c)}{=} \alpha [qq] + (1+\alpha) [0]\ .
\end{equation}
Similarly, we also have
\begin{equation}
\label{eqn:alpha-to-zero}
    1 [\alpha] \stackrel{(c)}{=} \alpha [q\rightarrow q] + (1-\alpha) [0]\ .
\end{equation}

\begin{figure}
    \centering
    \begin{tikzpicture}[scale=1.2]
        \draw (0,0) -- (1,1);
        \draw (0,0) -- (1,-1);
        \draw (1,1) -- (6,1); % B line
        \draw (1,-1) -- (2,-1); % first A line
        \draw (0,-1.5) -- (2,-1.5); % second A line
        \draw (2,-0.8) -- (3,-0.8) -- (3,-1.8) -- (2,-1.8) -- (2,-0.8); %U box
        \draw[thick] (3,-1) -- (5,1); %\alpha  
         \draw[->,thick] (3,-1) -- (4,0); %\alpha  
        \draw (3,-1.5) -- (6,-1.5); 
        \node[above] at (1.2,1) {$B$};
        \node[above] at (1.2,-1) {$A$};
        \node[above] at (0.2,-1.5) {$A$};
        \draw (6,1) -- (7,-0.25);
        \draw (6,-1.5) -- (7,-0.25);
        \node[below] at (0.2,-1.5) {$\alpha|0\rangle_A+\beta|1\rangle_A$};
        \node[right] at (2.3,-1.3) {$U_A$};
        \node[above] at (4,0.1) {$\alpha$};
        % \draw [decorate, decoration = {brace}] (6.2,1) --  (6.2,-1.5);
        \node[right] at (7.2,-0.3) {$\alpha|0\rangle_A|0\rangle_B+\beta|1\rangle_A |1\rangle_B$};
    \end{tikzpicture}
    \caption{Circuit diagram for converting $\alpha$-bits into cobits (the forward direction of \eqref{eqn:alpha-dense-coding}).}
    \label{fig:alpha-to-cobits}
\end{figure}
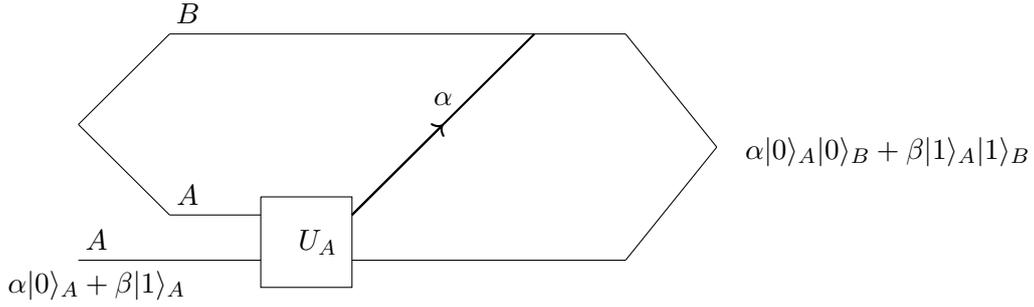
\section{$\alpha$-bit state merging with catalytic entanglement}
\label{sec:catalytic}

In this section, we show that there's a simple protocol that realizes state merging using only $\alpha$-bits.
\begin{theorem}
\label{thm:alpha-merge-cat}
For a state $|\psi\rangle_{ABR}$, the rate of $\alpha$-bits needed to achieve state merging is $\frac{I(A:R)_\psi}{1+\alpha}$ in the asymptotic limit when catalytic entanglement is allowed. That is, we have
\begin{equation}
\label{eqn:alpha-state-merging-catalytic-general}
   \langle \psi_{ABR}\rangle + \frac{I(A:R)_\psi}{1+\alpha} [\alpha] \stackrel{(c)}{\geq} \langle \psi_{A'BR} \rangle + (H(A)_\psi-\frac{I(A:R)_\psi}{1+\alpha})[qq]\ .
\end{equation}
This $\alpha$-bit rate is also optimal.

When $\frac{H(A|B)_\psi}{H(A)_\psi  }\leq \alpha$, no net entanglement is consumed, and when the equality holds,
\begin{equation}
\label{eqn:alpha-state-merging-catalytic}
  \langle \psi_{ABR}\rangle + H(A)_\psi [\alpha] \stackrel{(c)}{\geq} \langle \psi_{A'BR} \rangle.
\end{equation}
\end{theorem}

\begin{figure}
    \begin{tikzpicture}[scale=0.8]
        \draw (0,0) -- (12,0); % B line
        \draw (0,0) -- (1,1); 
        \draw (1,1) -- (14,1); % R line
        \draw (0,0) --(1,-3);
        \draw (1,-3) -- (1.5,-3);% A line
        \draw (1.5,-2.5) -- (2.5,-2.5) -- (2.5,-3.5) -- (1.5,-3.5) -- (1.5,-2.5); % U box
        \draw (2.5,-3) -- (3.5,-3); % line after U
        \draw (3.5,-2.5) -- (4.5,-2.5) -- (4.5,-3.5) -- (3.5,-3.5) -- (3.5,-2.5); % first U_A box
        \draw (2.5,-1.6) -- (3.,-0.6); %first EPR pair
        \draw (2.5,-1.6) -- (3,-2.6);
        \draw (3,-2.6) -- (3.5,-2.6);
        \draw (3,-0.6) -- (12,-0.6); %first EPR pair horizontal line
        \draw[thick] (4.5,-3) -- (6.5,-0.6); % first \alpha line
        \draw[->,thick] (4.5,-3) -- (5.5,-1.8);
        \node[above] at (5.5,-1.6) {$\alpha$};
        \draw (4.5,-3) -- (8.,-3); % line after first U_A
        \draw (7,-2) -- (7.3,-1.4); % second EPR pair
        \draw (7,-2) -- (7.3,-2.6);
        \draw (7.3,-2.6) -- (8,-2.6);
        \draw (8,-2.5) -- (9,-2.5) -- (9,-3.5) -- (8,-3.5) -- (8,-2.5); % second U_A box
        \draw (7.3,-1.4) -- (11,-1.4); % second EPR pair horizontal line
        \draw (11,-1) -- (11.8,-1) -- (11.8,-1.8) -- (11,-1.8) -- (11,-1); % X box
        \draw (11.4,-1) -- (11.4,-0.6);
        \draw[thick] (9,-3) -- (10.5,-1.4); % second \alpha line
        \draw[->, thick] (9,-3) -- (9.75,-2.2);
        \node[above] at (9.75, 
        -2.1) {$\alpha$};
        \draw (12,0.2) -- (13,0.2) -- (13,-0.8) -- (12,-0.8) -- (12,0.2); %V box
        \draw (11.8,-1.4) -- (12.2,-1.4);
        \draw (9,-3) -- (12.2,-3); % line after second U_A
        \draw (12.2,-3) -- (12.7,-2.2);
        \draw (12.2,-1.4) -- (12.7,-2.2);
        \draw (2.5,-3.5) -- (2.5,-4);
        \draw (2.5,-4) -- (14,-4); %bottom line
        \draw (13,-0.6) -- (14,-0.6);
        \draw (14,-4) -- (15,-2.3);
        \draw (14,-0.6) -- (15,-2.3);
        \draw (13,0) -- (14,0);
        \draw (14,0) -- (15,0.5);
        \draw (14,1) -- (15,0.5);

        \node[right] at (0.35,-2.8) {$A$};
        \node[above] at (1,0) {$B$};
        \node[above] at (1,1) {$R$};
        \node[right] at (1.7,-3) {$U$};
        \node[right] at (8.05,-3) {$U_A$};
        \node[right] at (3.55,-3) {$U_A$};
        \node[right] at (12.2,-0.25) {$V$};
        \node[right] at (11.1,-1.3) {$X$};
        \draw[black] plot [mark=*, mark size=2] coordinates{(11.4,-0.6)};
        \node[left] (0,0) {$|\psi\rangle_{ABR}$};
        \node[right] at (15,-2) {$|\Phi\rangle_{AB}$};
        \node[right] at (15,1) {$|\psi\rangle_{A'BR}$};
        \node[right] at (12.7,-2) {$|\Phi\rangle_{AB}$};
        \node[right] at (5.5,-2) {$|\Phi\rangle_{AB}$};
        \node[right] at (1,-1.5) {$|\Phi\rangle_{AB}$};
        \end{tikzpicture}
    \caption{Circuit diagram for $\alpha$-bit state merging with catalytic entanglement.}
    \label{fig:alpha-state-merging-cat}
\end{figure}
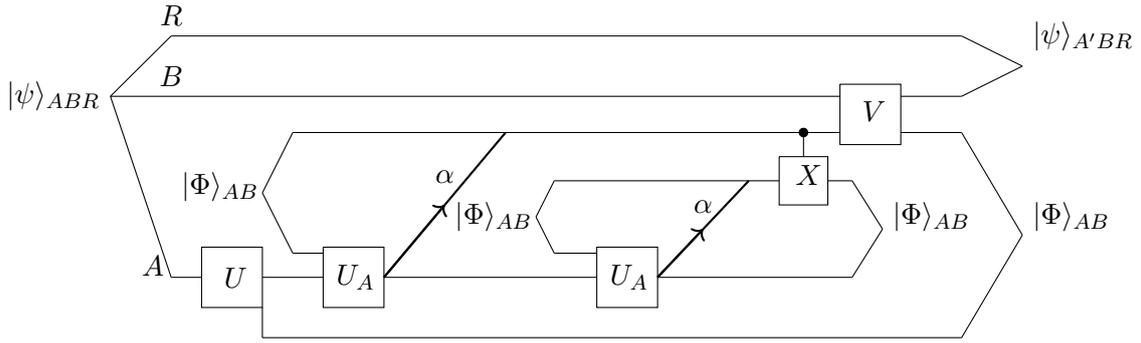
\subsection{Achievability}
First, we establish the achievability part. The idea is to simply combine the mother protocol, coherent teleportation, and $\alpha$-bit coherent dense coding. A circuit diagram that illustrates this protocol is in Fig.~\ref{fig:alpha-state-merging-cat}.

Starting with $\alpha$-bits, we first borrow a corresponding number of ebits to obtain cobits using \eqref{eqn:alpha-dense-coding}. Then we can use the cobits to do teleportation using \eqref{eqn:coherent-dense-coding}. In this step, we can also obtain extra ebits that we need to return. Lastly, we can perform the mother protocol with the qubits that we obtain in the last step. Combining
\eqref{eqn:fqsw}, \eqref{eqn:coherent-dense-coding}, and \eqref{eqn:alpha-dense-coding} after multiplying with appropriate values we can obtain
\begin{equation}
\label{eqn:alpha-state-merging-catalytic-general-before}
   \langle \psi_{ABR}\rangle + \frac{I(A:R)_\psi}{1+\alpha} [\alpha] + \frac{I(A:R)_\psi}{1+\alpha} [qq] \geq \langle \psi_{A'BR} \rangle + H(A)_\psi[qq]\ .
\end{equation}
Having a catalytic protocol means that we can combine the ebits to the RHS and obtain

\begin{equation}
% \label{eqn:alpha-state-merging-catalytic-general}
   \langle \psi_{ABR}\rangle + \frac{I(A:R)_\psi}{1+\alpha} [\alpha] \stackrel{(c)}{\geq} \langle \psi_{A'BR} \rangle + (H(A)_\psi-\frac{I(A:R)_\psi}{1+\alpha})[qq]\ .
\end{equation}

Since at the end we only want $\alpha$-bits as a resource to start with (up to the catalytic ebits), we require $H(A)_\psi-\frac{I(A:R)_\psi}{1+\alpha}\geq 0$, which means we need  $\alpha \geq \frac{H(A|B)_\psi}{H(A)_\psi}$. (Otherwise, the coefficient of [qq] on the RHS would be negative and implies that we need to consume ebits.) This is trivially satisfied if $H(A|B)_\psi<0$ but, in general, there is a nontrivial bound on the value of $\alpha$.

In the case when the equality holds, $\alpha = \frac{H(A|B)_\psi}{H(A)_\psi}$, which means $\frac{I(A:R)_\psi}{1+\alpha}=H(A)_\psi$ so the above equation reduces to \eqref{eqn:alpha-state-merging-catalytic}. 

The protocol is catalytic since we need to borrow ebits at the beginning to start the protocol. The rate of catalytic ebits used is $\frac{I(A:R)_\psi}{1+\alpha}=H(A)_\psi$.

Several variations of $\alpha$-bit state merging with catalytic entanglement have been studied~\cite{HP2017,AP2020}. Here, we make some distinctions.
In~\cite{HP2017}, the state merging is only written in terms of zero-bits.

\begin{equation}
    \langle \psi_{ABR}\rangle + I(A:R)_\psi [0] \stackrel{(c)}{\geq} \langle \psi_{A'BR}\rangle + I_c(A\rangle B)_\psi [qq]\ ,
\end{equation}
where $I_c(A\rangle B)_\psi:=-H(A|B)_\psi$ is the coherent information. This follows directly from the mother protocol \eqref{eqn:fqsw} and zero-bit teleportation \eqref{eqn:zero-bit-teleportation}.

In~\cite{AP2020}, the resource inequality is written in terms of qubits and zero-bits. In the i.i.d. limit this can be written as
\begin{equation}
    \langle \psi_{ABR}\rangle + H(A|B)_\psi [q\rightarrow q] + I(A:B)_\psi [0] \stackrel{(c)}{\geq} \langle \psi_{A'BR}\rangle [qq]\ .
\end{equation}
This is related to our \eqref{eqn:alpha-state-merging-catalytic} by setting $\alpha = \frac{H(A|B)_\psi}{H(A)_\psi}$ and using \eqref{eqn:alpha-to-zero}.

\subsection{Optimality}
We show that \eqref{eqn:alpha-state-merging-catalytic} is optimal. That is, we cannot perform state merging with less than $H(A)_\psi$ $\alpha$-bits.

First, we establish the optimality of the mother protocol: we need to send at least $I(A:R)_\psi/2$ qubits to achieve state merging. This comes from the fact that every qubit can carry at most two bits of information. Thus, starting with $I(A:R)_\psi$ bits of mutual information between $A$ and $R$ in the state $|\psi\rangle_{ABR}$, we need to send at least $I(A:R)_\psi/2$ qubits to destroy all the correlation between $A$ and $R$~\cite{ADHW2009mother}. Since decoupling is necessary for state merging, this establishes the optimality. 

The optimality of the catalytic $\alpha$-bit state merging then follows from the optimality of the mother protocol. Assume for the sake of contradiction that we can do $\alpha$-bit state merging with only $H(A)_\psi-\epsilon$ $\alpha$-bits for some $\epsilon>0$:
\begin{equation}
% \label{eqn:alphabit_state_merging}
  \langle \psi_{ABR}\rangle + (H(A)_\psi-\epsilon) [\alpha] \geq \langle \psi_{A'BR} \rangle \ .
\end{equation}
Then, since \eqref{eqn:coherent-dense-coding},\eqref{eqn:alpha-dense-coding} are both equalities, we can run the protocols in reverse and obtain the following resource inequality:
\begin{equation}
% \label{eqn:fqsw}
\langle \psi_{ABR}\rangle + \frac{1}{2}I(A:R)_\psi 
\left(\frac{H(A)_\psi-\epsilon}{H(A)_\psi}\right)[q\rightarrow q] \geq  \frac{1}{2}I(A:B)_\psi\left(\frac{H(A)_\psi-\epsilon}{H(A)_\psi}\right) [qq]  + \langle \psi_{A'BR} \rangle\ .
\end{equation}

That is, we would be able to perform the mother protocol with less than $I(A:R)_\psi/2$ qubits! This leads to a contradiction and thus establishes the optimality of \eqref{eqn:alpha-state-merging-catalytic}.

\section{$\alpha$-bit state merging without catalytic entanglement}
\label{sec:non-catalytic}
In the above section, we have shown how to perform state merging with available catalytic entanglement. It turns out that, in the absence of this catalytic entanglement, one needs strictly \textit{more} $\alpha$-bits to achieve state merging for all $\alpha <1$. 

\begin{theorem}\label{thm:noncatalytic}
For a state $|\psi\rangle_{ABR}$, the rate of $\alpha$-bits to achieve state merging while the use of catalytic entanglement is prohibited is $\frac{I(A:R)_\psi}{2\alpha}$ in the asymptotic limit. That is, the following resource inequality is optimal:
\begin{equation}
\label{eqn:alpha-bit-non-catalytic}
      \langle \psi_{ABR}\rangle + \frac{I(A:R)_\psi}{2\alpha} [\alpha] \geq \langle \psi_{A'BR} \rangle + \frac{I(A:B)_\psi}{2}[qq]\ .
\end{equation}
\end{theorem}

First, note that if $\alpha=1$, then this is exactly the mother protocol in~\eqref{eqn:fqsw}, and the catalytic and non-catalytic rates match. Generally when $\alpha<1$, the required $\alpha$-bits rate in the non-catalytic approach~\eqref{eqn:alpha-bit-non-catalytic} is always more than in the catalytic approach~\eqref{eqn:alpha-state-merging-catalytic-general}, which requires an $alpha$-bit rate of only $\frac{I(A:R)_\psi}{1+\alpha}$ as $\alpha \leq 1$. On the other hand, the difference in $\alpha$-bit consumption also gives rise to more ebits produced in the non-catalytic approach. See Fig.~\ref{fig:compare-rate} for a direct comparison. We also note that~\eqref{eqn:alpha-bit-non-catalytic} implies that zero-bits alone cannot achieve non-catalytic state merging, as the rate becomes divergent.

\subsection{Achievability}
\label{sec:non-catalytic-protocol}
The protocol is sketched as follows. (We also include a more detailed argument below. 
we begin with a sketch of the protocol before giving a more detailed argument. Throughout, we omit sublinear $o(n)$ corrections when specifying the dimensions of various subspaces. We start with a state $|\psi\rangle_{ABR}$. We begin $n$ copies of the state $|\psi\rangle_{ABR}$. First, we apply a Haar random on $A$ and divide the resulting state into two subsystems $C$ and $A'$. From the proof of the mother protocol, we know that the minimal rate of $|C|$ for $A'$ to decouple from $R$ is $\frac{1}{n}\log{|C|}=\frac{1}{2}I(A:R)_\psi$. 

Next, we append an auxiliary state $|00...0\rangle_{C'}$ to $C$ so that the total dimension of $CC'$ is roughly $2^{\frac{n I(A:R)_\psi}{2\alpha}}$. After doing this, the state on $C$ is now a subspace with dimension $2^{I(A:R)_\psi/2}$ in a space with dimension $2^{\frac{I(A:R)_\psi}{2\alpha}}$. 

We can now send $CC'$ to $B'$ as $\alpha$-bits. From the definition of $\alpha$-bits, $B'$ can recover $CC'$ since $2^{I(A:R)_\psi/2} \leq 2^{\frac{I(A:R)_\psi}{2\alpha}\alpha}$. Since Bob can recover the state on $CC'$, the environment of the channel decouples from the reference and gives rise to extra entanglement. In the end, Uhlmann's theorem implies there is a decoding isometry $D$ on $B$ that concludes the protocol.\\

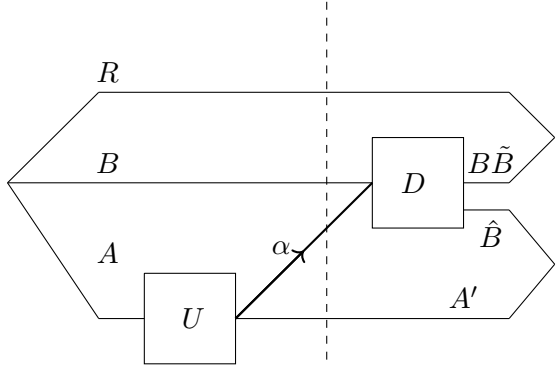
\begin{figure}\centering

    \centering {
    \begin{tikzpicture}[scale=1.2]
        \draw (0,0) -- (4,0); % B line
        \draw (0,0) -- (1,-1.5); 
        \draw (0,0) -- (1,1);
        \draw (1,-1.5) -- (1.5,-1.5); % A line
        \draw (1,1) -- (5.5,1); % R line
        \draw (5,0) -- (5.5,0); %B\tilde{B} line
        \draw (5.5,1) -- (6,0.5);
        \draw (5.5,0) -- (6,0.5);
        \draw (5,-0.3) -- (5.5,-0.3); %\hat{B} line
        \draw (2.5,-1.5) -- (5.5,-1.5); % A' line
        \draw (5.5,-0.3) -- (6,-0.9);
        \draw (5.5, -1.5) -- (6,-0.9);
        \draw (1.5,-1) -- (1.5,-2) -- (2.5,-2) -- (2.5, -1) -- (1.5, -1); %U box
        \draw (4,0.5) -- (4,-0.5) -- (5,-0.5) -- (5,0.5) -- (4,0.5); % D box
        \draw[thick] (2.5,-1.5) -- (4,0); % alpha line
        \draw[->,thick] (2.5,-1.5) -- (3.25,-0.75); % alpha line
        \draw[dashed] (3.5,2) -- (3.5,-2); %dahsed line
        \node[right] at (1.8,-1.5) {$U$};
        \node[above] at (1.1,1) {$R$};
        \node[above] at (1.1,0) {$B$};
        \node[above] at (1.1,-1) {$A$};
        \node[above] at (3,-0.9) {$\alpha$};
        \node[above] at (5,-1.5) {$A'$};
        \node[right] at (4.2,0) {$D$};
        \node[above] at (5.3,0) {$B\tilde{B}$};
        \node[below] at (5.3,-0.3) {$\hat{B}$};
    \end{tikzpicture}
    }
    \caption{Circuit diagram for the $\alpha$-bit state merging protocol without catalytic entanglement. The line with the $\alpha$ indicates the $\alpha$-bit channel in Fig.~\ref{fig:alpha}. $U$ is the Haar random applied by Alice and $D$ is the final isometry used by Bob to recover the state.}
\label{fig:circ-no-catalytic}
\end{figure}

% This constraint on $C$ can be rewritten as a constraint on $\alpha$: 
% \begin{equation}
% \label{eqn:alpha-constraint-noncatalytic}
%  k\alpha \geq \frac{1}{2}I(A:R)_\psi
% \end{equation}
% which is the achievability result we want to show.

% \begin{figure}
%     \centering
%     \includegraphics[width=5cm]{figs/alpha-amt-compare.jpeg}
%     \caption{}
%     \label{fig:alpha-amt-compare}
% \end{figure}

The following is a more detailed description of the procedure. See Fig.~\ref{fig:circ-no-catalytic} for the labeling.

1. The initial state is $|\psi\rangle_{ABR}^{\otimes n}$, $n$ copies of $|\psi\rangle_{ABR}$. Alice performs a Schumacher compression on $A^n$, projecting her share to its typical subspace, $A_S$. Then, she divided the $A_S$ system into two subsystems $A'$ and $C$ with $\log{|C|}=\frac{n}{2}I(A:R)_\psi$ and $\log{|A'|}=\frac{n}{2}I(A:B)_\psi$, omitting sublinear corrections as noted earlier. Alice then applies a Haar random unitary on $A_S$. From the proof of the mother protocol, we know that $A'$ is decoupled from $R$. Now we need to send $C$ to Bob with $\alpha$-bits.

2. Append some auxiliary state $C'$ so that $\log{|CC'|}=\frac{n I(A:R)_\psi}{2\alpha}$:
Therefore, by construction,  $C$ is in a subspace of $A_S$ with $|C|=\frac{n I(A:R)_\psi}{2}$ (the subspace that is spanned by the first $\frac{n}{2}I(A:R)_\psi$ qubits of $A$).

3. Alice sends the state on $CC'$ to Bob with $\alpha$-bits through the given $\alpha$-bit channel with Stinespring dilation $U_{\alpha}:CC'\rightarrow B'E$. Bob has a recovery channel with Stinespring dilation $V: B'\rightarrow CC'E'$ such that 
\begin{equation}
    (V \otimes \operatorname{Id}_E)U_{\alpha}|\chi\rangle_{CC'} \approx |\chi\rangle_{CC'} \otimes |\Phi \rangle_{EE'}
\end{equation}
for any state $|\chi\rangle_{CC'} \in C$.

This holds since we have the $\alpha$-bit channel with rate $\frac{I(A:R)_\psi}{2\alpha}$, and
\begin{equation}
    2^{\frac{1}{2}I(A:R)_\psi} \leq 2^{\frac{I(A:R)_\psi}{2\alpha}\alpha}\ .
\end{equation}

Bob now holds $CC'B$ and Alice holds $A'$.

4. From step 1, we know that $R$ and $A'$ are decoupled. Bob applies the isometry $D$ implied by Uhlmann's theorem and the final state is $|\psi\rangle_{B\tilde{B}R}^{\otimes n}\otimes |\Phi\rangle_{A'\hat{B}}^{\otimes n}$. This is achieved with error $\epsilon \rightarrow 0$ for sufficiently large $n$.

\subsection{Optimality}

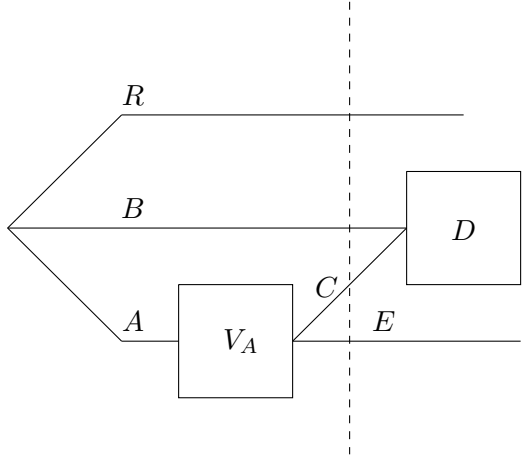
\begin{figure}\centering
    \centering {
    \begin{tikzpicture}[scale=1.5]
        \draw (0,0) -- (3.5,0);
        \draw (0,0) -- (1,-1);
        \draw (0,0) -- (1,1);
        \draw (1,-1) -- (1.5,-1);
        \draw (1,1) -- (4,1);
        \draw (1.5,-0.5) -- (1.5,-1.5) -- (2.5,-1.5) -- (2.5, -0.5) -- (1.5, -0.5);
        \draw (2.5,-1) -- (4.5,-1);
        \draw (3.5,0.5) -- (3.5,-0.5) -- (4.5,-0.5) -- (4.5,0.5) -- (3.5,0.5);
        \draw (2.5,-1) -- (3.5,0);
        \draw[dashed] (3,2) -- (3,-2);
        \node[right] at (1.8,-1) {$V_A$};
        \node[above] at (1.1,1) {$R$};
        \node[above] at (1.1,0) {$B$};
        \node[above] at (1.1,-1) {$A$};
        \node[above] at (2.8,-0.7) {$C$};
        \node[above] at (3.3,-1) {$E$};
        \node[above] at (4,-0.2) {$D$};
    \end{tikzpicture}
    }
    \caption{Circuit diagram for state merging starting with some initial state $|\psi\rangle_{ABR}$. $B$ can achieve state merging if and only if $R$ and $E$ decouple at the dashed line. In the context of mother protocol, this can be realized by choosing $V_A$ as Haar random. In the context of proving the optimality of $\alpha$-bit state merging without catalytic entanglement, the box $V_A$ represents all possible isometries $A$ can perform, including the encoding and the Stinespring dilation of the $\alpha$-bit channel.}
\label{fig:non-catalytic-optimal-proof}
\end{figure}

We want to show that $\frac{I(A:R)_\psi}{2\alpha}$ is optimal for the $\alpha$-bit state merging \eqref{eqn:alpha-bit-non-catalytic}. Consider a setup as in Fig.~\ref{fig:non-catalytic-optimal-proof}. The box $V_A$ denotes all the isometry $A$ can perform on her end, which also includes the encoding and the dilation of the $\alpha$-bit channel. Line C represents the subsystem sent to B through the $\alpha$-bit channel, and line E represents the subsystem remaining at A. The goal is to show we need to send at least $\frac{I(A:R)_\psi}{2\alpha}$ $\alpha$-bits for any possible~$V_A$. 

Let $k$ be the required rate of $\alpha$-bits, so that our goal is to show $k\geq \frac{I(A:R)_\psi}{2\alpha}$. From the definition of $\alpha$-bits and theorem \ref{thm:subspace-decoupling}, we have
$$ ||\mathcal{N}^c-\mathcal{R}||_{\diamond}^{(2^{k\alpha})}
 \leq \epsilon\ , $$ 
where $\mathcal{N}$ is the $\alpha$-bit channel.

And since we only have $k$ $\alpha$-bits and not more, we have for any $\log |R| > k\alpha$,
$$ ||\mathcal{N}^c-\mathcal{R}||_{\diamond}^{|R|}
 > \epsilon\ .$$

From the definition of the diamond norm, this implies that for any $\log |R| > k\alpha$, there exists some $\rho^{AR}$ such that 
\begin{equation}
    ||\mathcal{N}^c(\rho^{AR})-\mathcal{R}(\rho^{AR}) ||_1 = ||\rho^{ER}-\pi^E\otimes \rho^{R} ||_1 > \epsilon\ .
\end{equation} 

In other words, we can always find some $|\psi\rangle_{ABR}$ so that $R$ and $E$ are not decoupled after the channel, which further implies $B$ cannot achieve state merging.

Thus, we need $\log |R| \le k\alpha$ to achieve state merging in the worst case (\eqref{eqn:alpha-bit-non-catalytic} should hold for all $|\psi_{ABR}\rangle$). Since $\log{|R|} \geq \frac{I(A:R)_\psi}{2}$, we conclude that $k \geq \frac{I(A:R)_\psi}{2\alpha}$ is necessary and this completes the proof of optimality.

\subsection{Does catalysis help?}\label{sec:randomisometry}

We have shown that the catalytic protocol can be more economical in $\alpha$-bit consumption than the non-catalytic protocol. Does this mean catalysis can generally help with state merging? 

It is important to emphasize that the catalytic protocol may consume ebits depending on the entropies of the state-to-be-merged in consideration. Hence, to make a fair comparison between them, we need to restrict to a subset of states such that the catalytic protocol doesn't consume ebits, just like in the non-catalytic setting. We denote this subset of pure states satisfying $H(A|B)_\psi\le\alpha H(A)_\psi$ as $\mathcal{K}_\alpha$.

If we restrict the merging task to $\mathcal{K}_\alpha$, the converse argument for the optimality of the non-catalytic protocol no longer applies because that is a worst-case converse for merging all states. Hence, the question of whether catalysis can help with $\alpha$-bit state merging hinges on whether one can strengthen the converse for $\mathcal{K}_\alpha$.

While it is possible that there is an alternative non-catalytic protocol that consumes fewer $\alpha$-bits to merge states in $\mathcal{K}_\alpha$, we \emph{conjecture} that our converse can be strengthened. That is to say, if $R\le I(A:R)_\psi/2\alpha$, there exists some channel that could send $R$ $\alpha$-bits per channel use but would fail to merge some state in $\mathcal{K}_\alpha$. It's unlikely that there is some state outside the $d^\alpha$-dimensional subspace that can be decoupled for all $\alpha$-bit channels.\\ %\JW{Evidence/arguments for this conjecture? What's a better way to phrase our main result?}

Furthermore, it is also possible to bypass our non-catalytic converse by restricting to specific $\alpha$-bit channels.  It is possible because the general non-catalytic protocol uses an $\alpha$-bit channel as a black box. If one uses a particular $\alpha$-bit channel in a white-box scenario, one could hope to utilize features of the channel to achieve state merging more efficiently.

In fact, we do have such an example. Consider the prototypical example of an $\alpha$-bit channel~\cite{HP2017,HP2018}, which is a typical random isometry $V_{A\to BE}$ followed by tracing out less than half of the output system $\tr_E$, ($|E|<|B|$). One can easily show that this noisy channel $\mathcal{N}_\alpha:\rho_A\mapsto \tr_E V_{A\to BE}\ \rho_A V_{A\to BE}^\dagger$ is can send an $\alpha$-dit with $\alpha:=\frac{\log|B|-\log|E|}{\log|A|}$ and $d=|A|$ with a small error. In the asymptotic limit, $\mathcal{N}_\alpha$ is an $\alpha$-bit channel that can send $\log |A|$  $\left(\frac{\log|B|-\log|E|}{\log|A|}\right)$-bits.

It is straightforward to show that for states in $\mathcal{K}_\alpha$, the channel $\mathcal{N}_\alpha$ can achieve non-catalytic state merging that consumes $H(A)_\psi$ of $\alpha$-bits. To see this, we observe that for any state $\psi_{ABR}$ in $\mathcal{K}_\alpha$ satisfies $H(A|B)_\psi\le \log|B|-\log|E|$, which guarantees the decoupling between $E$ and $R$ and hence a successful state merging. 

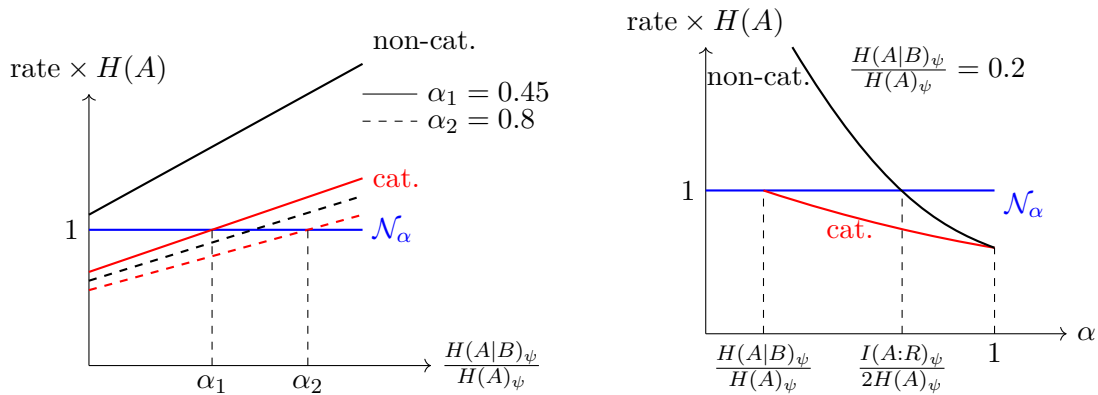
\begin{figure}
    \centering
    \begin{tikzpicture}[scale=1.8]
        \draw[->] (0,0) -- (2.5,0);
        \draw[->] (0,0) -- (0,2);
        \draw[thick, blue] (0,1) -- (2,1); % RI
        \draw[thick] (0,1/0.9) -- (2,2/0.9); % noncat alpha=0.45
        % \draw[thick] (0,1/1.8) -- (1,2/1.8);
        \draw[thick,red] (0,1/1.45) -- (2,2/1.45); % cat alpha=0.45
        \draw[thick,dashed] (0,1/1.6) -- (2,2/1.6); % noncat alpha=0.8
        \draw[thick,red,dashed] (0,1/1.8) -- (2,2/1.8); % cat alpha=0.8
        \node[right] at (2.5,0) {$\frac{H(A|B)_{\psi}}{H(A)_{\psi}}$};
        \node[above] at (0,2) {$\text{rate}\times H(A)$};
        \node[left] at (0,1) {$1$};
        \node[right,blue] at (2,1) {$\mathcal{N}_\alpha$};
        \node[right] at (2,2.4) {$\text{non-cat.}$};
        \node[right,red] at (2,2/1.45) {$\text{cat.}$};
        \node[below] at (0.9,0) {$\alpha_1$};
        \node[below] at (1.6,0) {$\alpha_2$};
        \draw[dashed] (0.9,0) -- (0.9,1);
        \draw[dashed] (1.6,0) -- (1.6,1);
        %add legend
        \node[right] at (2.4,2) {$\alpha_1=0.45$};
        \draw (2,2) -- (2.4,2);
        \node[right] at (2.4,1.8) {$\alpha_2=0.8$};
        \draw[dashed] (2.,1.8) -- (2.4,1.8);
    \end{tikzpicture}
    \hspace{0.5cm}
    \begin{tikzpicture}[scale=1.9]
        \draw[->] (0,0) -- (2.5,0);
        \draw[->] (0,0) -- (0,2);
        \draw[thick, blue] (0,1) -- (2,1); % RI
        \draw[thick,red] (0.4, 1) .. controls (1,1.2/1.5) and (1.6,1.2/1.8) ..  (2., 1.2/2); %cat
        \draw[thick] (0.6, 1.2/0.6) .. controls (1.2,1) and (1.6, 1.2/1.6)..  (2, 1.2/2); %non-cat
        \node[right] at (2.5,0) {$\alpha$};
        \node[above] at (0,2) {$\text{rate}\times H(A)$};
        \node[left] at (0,1) {$1$};
        \node[right,blue] at (2,0.9) {$\mathcal{N}_\alpha$};
        \node[above,red] at (1,0.6) {$\text{cat.}$};
        \node[left] at (0.8,1.8) {$\text{non-cat.}$};
        \node[above] at (1.6,1.6) {$\frac{H(A|B)_{\psi}}{H(A)_{\psi}}=0.2$};
        \draw[dashed] (0.4,0) --(0.4,1);
        \node[below] at (0.4,0) {$\frac{H(A|B)_{\psi}}{H(A)_{\psi}}$};
        \draw[dashed] (2,0) -- (2,1.2/2);
        \node[below] at (2,0) {$1$};
        \draw[dashed] (1.36,0) --(1.36,1);
        \node[below] at (1.36,0) {$\frac{I(A:R)_{\psi}}{2H(A)_{\psi}}$};

    \end{tikzpicture}
    % \begin{tikzpicture}[scale=1.9]
    %     \draw[->] (0,0) -- (2,0);
    %     \draw[->] (0,0) -- (0,2);
    %     \draw[thick, blue] (0,1) -- (2,1); % RI
    %     \draw[thick] (0,1/1.6) -- (2,2/1.6); % noncat alpha=0.8
    %     \draw[thick,red] (0,1/1.8) -- (2,2/1.8); % cat alpha=0.8
    %     \node[right] at (2,0) {$\frac{H(A|B)_{\psi}}{H(A)_{\psi}}$};
    %     \node[above] at (0,2) {$\text{rate}\times H(A)$};
    %     \node[left] at (0,1) {$1$};
    %     \node[right,blue] at (2,0.9) {$\mathcal{N}_\alpha$};
    %     \node[right] at (2,1.42) {$\text{non-cat.}$};
    %     \node[right,red] at (2,1.2) {$\text{cat.}$};
    %     \node[above] at (0.6,1.7) {$\alpha=0.8$};
    %     \draw[dashed] (1.6,0) -- (1.6,1);
    %     \node[below] at (1.6,0) {$\alpha$};
    % \end{tikzpicture}
    \caption{(Left):The rate of $\alpha$-bit state merging as a function of $\frac{H(A|B)_{\psi}}{H(A)_{\psi}}$, using three different protocols: catalytic protocol in Sec.~\ref{sec:catalytic} (red), non-catalytic protocol in Sec.~\ref{sec:non-catalytic} (black), and the prototypical random isometry channel $\mathcal{N}_\alpha$ discussed in Sec.~\ref{sec:randomisometry} (blue). The solid lines show the rate using the $\alpha$-bit channel with $\alpha=0.45$, and the dashed lines show the rate with $\alpha=0.8$. The catalytic rate is the lowest, and it coincides with the non-catalytic protocol done with $\mathcal{N}_\alpha$ when $\alpha=\frac{H(A|B)_{\psi}}{H(A)_{\psi}}$. (Right): The rate of $\alpha$-bit state merging as function of~$\alpha$, while fixing $\frac{H(A|B)_{\psi}}{H(A)_{\psi}}=0.2$. The non-catalytic rate diverges at $\alpha=0$. The catalytic rate coincides with the non-catalytic rate when $\alpha$=1. For the subset of states in $\mathcal{K}_\alpha$ satisfying $\frac{I(A:R)_\psi}{2H(A)_{\psi}}\le\alpha$, the protocol with $\mathcal{N}_\alpha$ is the least efficient.}
    \label{fig:compare-rate}
\end{figure}

We can now compare $H(A)_\psi$ with the catalytic rate $I(A:R)_\psi/(1+\alpha)$ and the non-catalytic rate $I(A:R)_\psi/(2\alpha)$. This is shown in Fig.~\ref{fig:compare-rate}. Since $H(A|B)_\psi\le \alpha H(A)_\psi$ for $\psi\in\mathcal{K}_\alpha$, we have 
\begin{equation}
H(A)_\psi\ge \frac{I(A:R)_\psi}{1+\alpha}    
\end{equation}
with the equality achieved for states exactly satisfying  $\frac{H(A|B)_\psi}{H(A)_\psi}=\alpha$. 

Furthermore, if $\frac{H(A|B)_\psi}{H(A)_\psi}\le 2\alpha-1$, i.e., $ \frac{I(A:R)_\psi}{2H(A)_{\psi}}\le\alpha$, which is only possible when~$\alpha\ge1/2$, then 
\begin{equation}
    H(A)_\psi\ge\frac{I(A:R)_\psi}{2\alpha}\ge \frac{I(A:R)_\psi}{1+\alpha}\ ,
\end{equation}
which means this protocol is even worse than the generic non-catalytic rate in Theorem \ref{thm:noncatalytic}. 

Hence, we conclude that the channel $\mathcal{N}$ can achieve non-catalytic state merging for states satisfying $\frac{H(A|B)_\psi}{H(A)_\psi}\le\alpha$, and it generally consumes more $\alpha$-bits than the optimal catalytic rate. For states satisfying $\frac{H(A|B)_\psi}{H(A)_\psi}=\alpha$, the prototypical example of $\alpha$-bit channel does achieve the optimal state merging rate even in the non-catalytic setting. 

On the other hand, for states satisfying $\frac{I(A:R)_\psi}{2H(A)_{\psi}}\le\alpha$, it has worse efficiency compared to the general non-catalytic protocol. This is because the merging task is relatively too easy for the channel and we are not utilizing the full communication capacity of it. 

As we shall see next, this non-catalytic protocol $\mathcal{N}_\alpha$ is very much relevant to the mechanism of Entanglement Wedge Reconstruction in AdS/CFT.

\section{Implications for entanglement wedge reconstruction}
\label{sec:holography}

As alluded to in the introduction, $\alpha$-bit state merging has found application in entanglement wedge reconstruction (EWR) in the context of AdS/CFT correspondence, as first proposed by Akers-Penington (AP) in~\cite{AP2020}. In this section, we elaborate on this connection and discuss some subtleties the appear in light of our results. %We first briefly review the bare minimum background and then discuss the implications of our results. 

\subsection{EWR in AdS/CFT}
AdS/CFT correspondence is a proposed duality between string theory in AdS$\times S^5$ spacetime and $\mathcal{N}=4$ super Yang-Mills theory on the boundary~\cite{Maldacena1997adscft}. Numerous studies have provided strong evidence for this duality~\cite{Wit1998AdSHolography, BanDouHor1998AdSdynamics,HeePenPol2009Holography, AGMOO2000review}. In particular, by studying the so-called AdS/CFT dictionary, it is found that this correspondence holds at the subregion level~\cite{hkll,BouFreLei2012null,BouLeiRos2012lightsheet,CzeKarRam2012gravitydual}. That is, a subregion $\Sigma$ of the boundary theory is dual to a subregion $\sigma$ in the bulk spacetime, which is known as \emph{subregion-subregion duality}. Operationally, this means that the information in the bulk subregion can be reconstructed in the corresponding boundary subregion. A more detailed introduction to this subject can be found in~\cite{harlow2018tasi}.

A naive candidate for $\sigma$ is the causal wedge, which is the bulk region that has causal contact with $\Sigma$. The causal wedge can be explicitly reconstructed by the HKLL procedure using the equations of motion of the bulk theory~\cite{hkll}. The surprising fact is that in many cases the region $\sigma$, called the entanglement wedge of $\Sigma$, can be strictly larger than the causal wedge. The specific procedure used to reconstruct operators supported on this bulk region $\sigma$ from operators supported on the boundary region $\Sigma$ is called the \emph{entanglement wedge reconstruction}~\cite{JLMS2015,XiHarWal2016Rec,CotHayPen2019EWR}.
A closely related idea is the Ryu-Takayanagi surface, which is later generalized to the quantum extremal surface (QES)~\cite{RyuTak2006RT,HubRanTak2007HRT,LewMal2013LM, FauLewMal2013FLM,EngWal2015QES}. This is a prescription that equates the entropy of a boundary subregion to the area of some surface in the bulk.  Given a boundary subregion $\Sigma$, its QES $\gamma_\Sigma$ is the surface that minimizes the generalized entropy among all extremal surfaces:
\begin{equation}
\label{eqn:QES}
    S(\Sigma)= {\min \text{ext}}_{\gamma_\Sigma} \left[
    \frac{\text{Area}(\gamma_\Sigma)}{4 G_N} + S_{bulk} (\gamma_\Sigma)\right]\ ,
\end{equation}
where $S_{bulk} (\gamma_\Sigma) $ is the bulk entropy in the region enclosed by the QES, and ``ext'' means to extremize over all surfaces homologous to $\Sigma$. We will only discuss the time-symmetric spatial slices, so ``ext'' above can be omitted. 

\begin{figure}
    \centering
    {
    \begin{tikzpicture}[scale=0.7]
    \draw[thick] (0,0) circle (3);
    \draw[red, very thick] (2.12, 2.12) arc (45:-45:3);
    \draw[blue, very thick] (2.12, 2.12) to [out=-120,in=120] (2.12, -2.12);
    % \draw[blue, very thick] (2.12, 2.12) .. controls (-0.3,0.5) and (-0.3,-0.5) ..  (2.12, -2.12);
    \draw[blue, very thick] (2.12, 2.12) .. controls (-0.6,1.8) and (-0.6,-1.8) ..  (2.12, -2.12);
    \node[right,red] at (3,0) {$\Sigma$};
    \node[right] at (0.5,0) {$A$};
    \node[right] at (1.8,0) {$B$};
    \node[left,blue] at (1.5,2.1) {$\gamma_1$};
    \node[left,blue] at (2.7,1.4) {$\gamma_2$};
    \end{tikzpicture}
    }
    \caption{Given a boundary subregion $\Sigma$, the goal of entanglement wedge reconstruction is to recover the state in the wedge $\sigma=A \cup B$ from $\Sigma$. $B$ represents the part of the wedge that is close to $\Sigma$ and can be reconstructed by the HKLL procedure, and $A$ represents the remaining wedge that can be reconstructed by the $\alpha$-bit state merging.}
    \label{fig:wedge}
\end{figure}
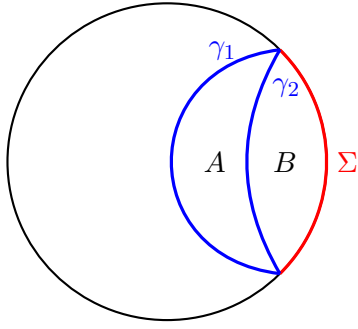

As an example, consider the setup in Fig.~\ref{fig:wedge}. There are two candidate QES surfaces $\gamma_1$ and $\gamma_2$ with areas $A_1$ and $A_2$, and with entanglement wedge $A \cup B$ and $B$, respectively. Suppose that we have 
\begin{equation}
    \frac{A_2}{4G_N} + H(B)\geq \frac{A_1}{4G_N} +H(AB)  \implies \frac{A_2-A_1}{4 G_N} \geq H(A|B)\ ,
\end{equation}
and then the entropy of $C$ is given by the generalized entropy $\frac{A_1}{4G_N} +H(AB)$. %This will be the case we are interested in.

Remarkably, the region enclosed by the QES is exactly the EW of $\Sigma$! In the above example, the entanglement wedge is therefore the larger region $\sigma= A\cup B$. This connection has been explained using the language of quantum error correction (QEC)~\cite{AlmDonHar2014QEC,HaPPY2015,Har2016RTQES,XiHarWal2016Rec}. The background geometry provides a bulk-to-boundary map, which can be effectively modeled by a (random) tensor network built out of the data of QES areas~\cite{Swingle_2012,HayNezQi2016RTN,Bao_2019,Akers_2019,AP2020}. We can view the bulk-to-boundary map as a QEC code that maps the logical degrees of freedom in the bulk to the boundary physical degrees of freedom. Then EWR only works if the erasure of the complementary boundary region is a recoverable error on the logical level.

\subsection{EWR as state merging}
Usually, we only demand that EWR work for a subspace of states, called the \emph{code subspace}. However, this is sometimes too restrictive.\emph{State-dependent} EWR only demands the task to work for operators acting on a particular state. In the Schrodinger picture description, EWR asks for a boundary state on the subregion that shares the same purification as the bulk state in the entanglement wedge. This is exactly the motivation for interpreting state-dependent EWR as state merging.

Let us consider the following simple setup to illustrate this idea, as shown in Fig.~\ref{fig:wedge}. Let $\Sigma$ be a boundary region held by Bob, and $B$ be the causal wedge of $\Sigma$. Since $B$ is the causal wedge, it is reconstructable from $\Sigma$ via the HKLL procedure, so Bob is also in control of $B$. Let $A$ be a bulk region held by Alice. Then we consider a bulk state $\psi_{AB}$ that is purified by some $\psi_{ABR}$, then the goal of EWR is to let Bob recover $\psi_{ABR}$ from $\Sigma$. It is sufficient to let Bob recover the state from the causal wedge $B$. This means that Alice can merge her share of the state with Bob.

There is an important conceptual distinction between EWR and state merging that is worth clarifying. In our setup, EWR is purely a ``spacelike'' task, where we try to identify the same quantum information residing in different spatial locations in the bulk, and no literal communication is involved. On the other hand, state merging is a ``timelike'' task, where Alice transfers some quantum information to Bob via communication. Nonetheless, there is no mathematical distinction between how we model the spacelike bulk-to-boundary dictionary/map and the timelike communication channel in state merging: They are both CPTP maps from Alice to Bob, $A\to B$. Hence, the mathematical results of state merging can be applied to study EWR, but we should bear in mind that EWR is strictly an unconventional ``spacelike-variant'' of state merging.

For state merging to work, Alice and Bob need to ``communicate'' something, which can be qubits, classical bits, or $\alpha$-bits. AP argued that there isn't enough ``bandwidth'' between Alice and Bob to perform conventional state merging with classical bits or qubits. Instead, one needs to resort to $\alpha$-bits, for which there is indeed enough bandwidth for the state merging. Namely, the bulk-to-boundary map sends enough $\alpha$-bits to make state merging possible for a subset of states. Their analysis essentially lands on the same catalytic protocol stated in our Theorem~\ref{thm:alpha-merge-cat},\footnote{\label{fn:AP}AP didn't mention $\alpha$-bit state merging nor the use of catalytic ebits. Instead, AP argued that state merging, or more generally the mother protocol, can be achieved by qubits plus zero-bits, which can be provided by the $\alpha$-bits. We think it is equivalent and more straightforward to state it as $\alpha$-bit state merging. Furthermore, AP's argument relating $\alpha$-bits, qubits, and zero-bits implicitly uses catalytic ebits, so eventually it would lead to the same result we have in Theorem~\ref{thm:alpha-merge-cat}.} which works most efficiently for those states in $\mathcal{K}_\alpha$.

In terms of our example in Fig.~\ref{fig:wedge}, the fact that $A$ can send enough $\alpha$-bits to $B$ is established in~\cite{HP2018}, where one of us showed that $\log |A|$ $\alpha$-bits can be sent with $\alpha=\frac{A_2-A_1}{4G_N \log |A|}$. It can help Alice and Bob (catalytically) merge typical states in $\mathcal{K}_\alpha$, satisfying $\frac{H(A|B)_\psi}{H(A)_\psi}\le\alpha$, such that $H(A)_\psi\approx \log |A|$ at the leading order. Therefore, for EWR to work on
these states, we require $\frac{A_2-A_1}{4G_N}\geq H(A|B)_\psi$. This indeed coincides with the entanglement wedge we found using the QES formula.

When we have multiple QES candidates for a disconnected boundary subregion, the bulk is divided into multiple subregions. Then the problem of identifying its EW becomes more complicated. The guiding principle is to identify how does quantum information flow across bulk subregions and eventually to the boundary. This is particularly relevant if we want to understand EWR from a purely bulk perspective, which can be generalized to a theory without a boundary dual \cite{Bousso2022wedges,Bousso2023holo,bousso2025simple,bousso2025fundamental,kaya2025hollow}. Hence, these considerations further motivate the idea of treating EWR as state merging among possibly multiple parties~\cite{multiparty,Colomer_2024}.\\

We should note that in~\cite{AP2020}, the analysis is performed in the one-shot setting. (See also~\cite{Akers2023oneshot}.) This is crucial because the EWR is meant to be a one-shot task; there are not multiple copies of the state available. For simplicity of presentation, we shall consider those typical states whose one-shot entropies are comparable to von Neumann entropies, such as $H(A)_\psi\approx \log |A|$. This is also the most common scenario studied in AdS/CFT. Hence, even though the rates we considered are presented as von Neumann entropies, they roughly coincide with the one-shot rates. For completeness, we also give the one-shot versions of our results in Appendix~\ref{sec:oneshot}.

\subsection{EWR as generic catalytic $\alpha$-bit state merging?}

We would like to raise the concern that it is unclear to what extent one can think of EWR as literally an $\alpha$-bit state merging process as AP described. In particular, the argument in~\cite{AP2020} implicitly makes use of the catalytic entanglement (cf. footnote~\ref{fn:AP}, but it is not discussed where one can obtain the catalysis from the bulk). 

One immediate solution to this is to borrow the vacuum entanglement in the bulk theory and use this to start the $\alpha$-bit state merging protocol. But how much entanglement do we need to borrow for catalysis?

In the asymptotic multi-shot setting, the usual way to get around this problem is to \emph{recycle} the catalytic ebits during the state merging process~\cite{DevHarWin2008resource}. Suppose that the protocol is run $n$ times, and each time we need to use $H(A)_\psi$ catalytic ebits. After each round, we can use the same ebits again. One can show that only $O(\log{n})$ catalytic ebits are required, leading to a sublinear consumption of catalytic entanglement, and thus a vanishing rate.

However, as AP argued, the task of EWR in holography is crucially a one-shot task, so one cannot recycle ebits. In that case, we show in Appendix~\ref{sec:oneshot} that the number of ebits needed in the one-shot version of our protocol is comparable to the number of $\alpha$-bits consumed, which corresponds to $O(1/G_N)$ in the gravity language.

If the EWR really functions as a one-shot catalyzed $\alpha$-bit state merging, one problem that immediately arises is that the manipulation of such a huge amount of entanglement would cause severe backreaction to the geometry. 

On the other hand, in the case where there is no access to catalytic ebits, we have shown that state merging generally requires more $\alpha$-bits than the catalytic version. Even for the smaller subset $\mathcal{K}_\alpha$, we believe that the optimal catalytic rate cannot be achieved non-catalytically. Therefore, one can only perhaps reconstruct a strictly smaller wedge if no catalytic ebits are allowed. We hence conclude it is unlikely that EWR operates as catalytic $\alpha$-bit state merging.

%The only problem is that since in each round, the ebits get degraded, meaning the ebits become only $\epsilon$ close to the perfect EPR pairs, we need to examine how much ebits we need to borrow at the beginning to ensure the error is still vanishing after all $n$ rounds. In the appendix, we show that only $O(\log{n})$ amount of catalytic ebits are required, leading to a sublinear consumption of catalytic entanglement. A similar argument can also be found in~\cite{DevHarWin2008resource}.

%We think that the issues that arise with catalytic entanglement do not affect the main conclusion of~\cite{AP2020}, in particular the refined QES formula.  However, one needs to be more careful when treating entanglement wedge reconstruction as a state merging process. More detailed formulations will be needed to take this idea further. 

\subsection{EWR as situation-specific $\alpha$-bit state merging}
\begin{figure}
    \centering
    {
    \begin{tikzpicture}[scale=0.7]
    \draw[thick] (0,0) circle (3);
    \draw[red, very thick] (2.12, 2.12) arc (45:-45:3);
    \draw[blue, very thick] (2.12, 2.12) to [out=-120,in=120] (2.12, -2.12);
    \draw[blue, very thick] (2.12, 2.12) .. controls (-0.6,1.8) and (-0.6,-1.8) ..  (2.12, -2.12);
    \node[right,red] at (3,0) {$\Sigma$};
    \draw[thick] (0.5,0.5) -- (1.3,0.5) -- (1.3,-0.3) -- (0.5,-0.3) -- (0.5,0.5); %V box
    \draw[thick] (-1.3,0.5) -- (-0.5,0.5) -- (-0.5,-0.3) -- (-1.3,-0.3) -- (-1.3,0.5); %left box
    \draw[thick] (-2,0.1) -- (-1.3,0.1); %leftmost line
    \draw[thick] (-0.7,-0.3) -- (-0.2,-1.2);
    \draw[thick] (1.3,0.1) -- (2.7,0.1); %rightmost line
    \draw[thick] (-0.5,0.1) -- (0.5,0.1); %between two boxes
    \draw[thick] (0.9,-0.3) -- (1.4,-1.2); %below V
    \node[right] at (0.5,-1.1) {$A$};
    \node[above] at (2.3,0.1) {$B$};
    \node[above] at (-0.2,0.1) {$E$};
    \node[right] at (0.5,0.1) {$V$};
    % \node[left,blue] at (1.2,1.6) {$\gamma_1$};
    % \node[left,blue] at (2.7,1.4) {$\gamma_2$};
    \end{tikzpicture}
    }
    \caption{Random tensor network as a bulk-to-boundary map.}
    \label{fig:wedge+RTN}
\end{figure}
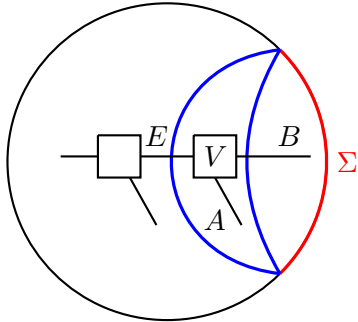
The remaining possibility is that EWR actually corresponds to a \emph{specific non-catalytic $\alpha$-bit state merging protocol} that utilizes particular features of the bulk-to-boundary map. This allows EWR to consume fewer $\alpha$-bits than the general non-catalytic protocol in Theorem~\ref{thm:noncatalytic}, such that it can send enough $\alpha$-bits to merge any state in $\mathcal{K}_\alpha$.

Recall that the bulk-to-boundary map between $A$ and $B$ can send $\log |A|$ of $\alpha$-bits with $\alpha=\frac{A_2-A_1}{4G_N \log |A|}$.  We have seen in Sec.~\ref{sec:randomisometry} that the random isometric channel $\mathcal{N}_\alpha$ has exactly this functionality of sending $\log |A|$ $\alpha$-bits, if the random isometry $V_{A\to BE}$ has the input size $\log|A|$, output size $\log|B|=A_2/4G_N$ and environment size $\log|E|=A_1/4G_N$. 

In connection to gravity, the bulk-to-boundary map can be modeled by a random tensor network where each bulk subregion is associated with a random isometry that maps the bulk degrees of freedom to adjacent boundary regions, see Fig.~\ref{fig:wedge+RTN}. This is also the model in which AP analyzed the EWR. The boundary-to-boundary map is then exactly modeled by a random isometric channel $\mathcal{N}_\alpha$. To the extent that the random tensor network is a good model for the bulk in the semiclassical regime, this is likely how gravity can manage to achieve state merging using the bulk-to-boundary map non-catalytically for states in~$\mathcal{K}_\alpha$.\\

It is interesting to note that the gravity protocol is generally less efficient than the catalytic protocol, where the catalytic ebits help to unlock the full communication bandwidth of the bulk-to-boundary channel. However, as for EWR, the upshot of being non-catalytic is more crucial for the reasons we discussed above. Moreover, for states satisfying $\frac{H(A|B)_\psi}{H(A)_\psi}\le 2\alpha-1$, the gravity protocol is also less efficient than the general non-catalytic protocol. This is reasonable because in this case the EWR is less demanding, and we are not fully utilizing the capability of the bulk-to-boundary channel.

\acknowledgements
We would like to thank Chris Akers and Geoff Penington for discussions, and Arjun Mirani for initial collaborations. The authors appreciate support from AFOSR (award FA9550-19-1-0369), DOE (Q-NEXT), CIFAR, and NSF (awards No. 2440805 and 2016245).

\appendix
\section{One-shot $\alpha$-bit state merging}\label{sec:oneshot}
Here we study the resource consumption of $\alpha$-bit state merging protocols done in one shot. We ask how big the dimension of an $\alpha$-dit that one needs to send in order to complete the merging task in one shot. For instance, the one-shot $\alpha$-bit dense coding says that for any dimension~$d$ and error tolerance $\delta>0$, there exists $\eps>0$ such that the task of sending $(1+\alpha)\log d$ cobits can be achieved by sending one $\alpha$-dit with error $\eps$ with the help $\log d$ ebits (cf. Theorem 5 in~\cite{HP2018}). Note that here the one-shot task is defined w.r.t. one copy of $\alpha$-dit. The resource inequality reads
\begin{equation}
   1[\alpha_d]_\varepsilon+\log d \, [qq] \ge (1+\alpha)\log d \, [q\to qq]_\delta
\end{equation}
where $[\alpha_d]_\varepsilon$ denotes an $\alpha$-dit with error $\varepsilon$, and $[q\to qq]_\delta$ denotes imperfect cobits with error $\delta$,\footnote{In the literature, the error of achieving the task is sometimes denoted as $\ge_\delta$ instead.} and we have $\delta(\varepsilon)\to 0$ as $\varepsilon\to 0$. 

%While the relevant resource is an $\alpha$-dit, we still want to count it as $\log d$ $\alpha$-bits so as to make the concatenation with other resources more conveniently. We hence divide $\log d$ on both sides to obtain the following resource inequality
%\begin{equation}
%   1[\alpha]_\varepsilon+[qq] \ge (1+\alpha) [q\to qq]_\delta
%\end{equation}
%and it is just a shorthand for the more precise resource inequality above.

Let's also revisit the one-shot mother protocol~\cite{datta2011apex}, that achieves the state transfer for a single copy of $\psi_{ABR}$. Let $\eps>0$, there exists a $\delta$-approximate state transfer mother protocol such that
\begin{multline}\label{eqn:oneshotmother}
    \langle \psi_{ABR}\rangle + \frac{1}{2}(H_{\max}^{\eps}(A)_\psi-H_{\min}^{\eps}(A|R)_\psi)[q\rightarrow q] \\ \geq  \frac{1}{2}(H_{\max}^\eps(A)_\psi+ H_{\min}^\eps(A|R)_\psi) [qq]  + \langle \psi_{A'BR} \rangle_\delta\ ,
\end{multline}
where we've omitted the negligible $\mathcal{O}(\log\eps)$ corrections to the rates, and $\delta(\eps)\to 0$ as $\eps\to 0$.

Let us consider the one-shot non-catalytic $\alpha$-bit state merging protocol. We can easily adapt our achievability argument given in the main text to the one-shot case, by changing the i.i.d. decoupling condition to the one-shot decoupling condition to its one-shot counterpart~\cite{dupuis2010decoupling}, that is we need $\log|C|=\frac{1}{2}[H_{\max}^{\eps}(A)_\psi-H_{\min}^{\eps}(A|R)_\psi]$. We hence obtain the resource inequality,
\begin{equation}
    \langle \psi_{ABR}\rangle + [\alpha_d]_\epsilon \geq  \frac{1}{2}(H_{\max}^\eps(A)_\psi+ H_{\min}^\eps(A|R)_\psi) [qq] + \langle \psi_{A'BR} \rangle_\delta\ ,
\end{equation}
where $\log d=\frac{1}{2\alpha}(H_{\max}^{\eps}(A)_\psi-H_{\min}^{\eps}(A|R)_\psi)$ and $\delta(\epsilon,\eps)\to 0$ as $\eps, \epsilon\to 0$.

It says that an $\alpha$-dit (with error $\epsilon$) of dimension $d=2^{\frac{1}{2\alpha}(H_{\max}^{\eps}(A)_\psi-H_{\min}^{\eps}(A|R)_\psi)}$ can be used to merge one copy of the state non-catalytically with an error $\delta(\eps,\epsilon)$.

Let's now consider the catalytic protocol. A key difference is that, in the one-shot setting, it's important to keep track of the catalysts because they can no longer be recycled as in the i.i.d. scenario. Hence, we don't allow \emph{subtracting} common ebit terms on both sides of the one-shot resource inequality.

Starting with the one-shot mother protocol~\eqref{eqn:oneshotmother}, we add more ebits to both sides and use coherent teleportation~\eqref{eqn:coherent-dense-coding} to obtain,
\begin{equation}
\begin{aligned}
&\langle \psi_{ABR}\rangle + (H_{\max}^{\eps}(A)_\psi-H_{\min}^{\eps}(A|R)_\psi)[q\rightarrow qq] \\\geq&
    \langle \psi_{ABR}\rangle + \frac{1}{2}(H_{\max}^{\eps}(A)_\psi-H_{\min}^{\eps}(A|R)_\psi)[q\rightarrow q]+\frac{1}{2}(H_{\max}^{\eps}(A)_\psi-H_{\min}^{\eps}(A|R)_\psi)[qq] \\\geq&  H_{\max}^\eps(A)_\psi [qq]  + \langle \psi_{A'BR} \rangle_\delta\ .
\end{aligned}
\end{equation}

Now using $\alpha$-bit dense coding, we have
\begin{multline}
    \langle \psi_{ABR}\rangle + [\alpha_d]_\epsilon+\frac{1}{(1+\alpha)}(H_{\max}^{\eps}(A)_\psi-H_{\min}^{\eps}(A|R)_\psi)[qq]\\\geq \langle \psi_{ABR}\rangle + (H_{\max}^{\eps}(A)_\psi-H_{\min}^{\eps}(A|R)_\psi)[q\rightarrow qq]_{\delta'} \geq  H_{\max}^\eps(A)_\psi [qq]  + \langle \psi_{A'BR} \rangle_{\Delta}\ ,
\end{multline}
where $\log d =(H_{\max}^{\eps}(A)_\psi-H_{\min}^{\eps}(A|R)_\psi)/(1+\alpha)$, $\delta'$ is the error on cobits from using $\alpha$-dit with error $\epsilon$, and the overall state merging error is denoted by $\Delta$, and $\Delta(\delta,\delta')\equiv\Delta(\eps,\epsilon)\to 0$ as $\eps,\epsilon\to 0$ . 

We see that we need a considerable amount of entanglement to \emph{activate} the protocol and we return some back as the yield. Since we cannot recycle the ebits here, the one-shot protocol is significantly more demanding than the asymptotic rate suggests, as the activation ebits are comparable in number to the $\alpha$-bits needed.

% \section{Error Analysis of catalytic ebits recycling}
% Consider a one-shot
% \label{app:recycle}

\bibliographystyle{utphys}
\bibliography{ref}
\end{document}